\documentclass[10pt]{article}
\usepackage{latexsym}
\usepackage{amssymb}
\usepackage{amsmath}
\usepackage{amscd}
\usepackage{amsthm}
\usepackage[left=2cm,top=2.5cm,right=2.5cm,bottom=1.5cm]{geometry}
\usepackage[dvips]{graphicx}
\usepackage{hyperref}

\begin{document}
\begin{center}
\Large{\bf{Some Plane Symmetric Inhomogeneous Cosmological Models in the Scalar-Tensor Theory of Gravitation}}

\vspace{10mm}

\normalsize{Ahmad T Ali$^{\dag, \$}$, Anil Kumar Yadav$^{\ddag}$ 
and S R Mahmoud$^{\dag, \S}$} \\

\vspace{4mm}
\normalsize{$^\dag$ King Abdul Aziz University,\\
Faculty of Science, Department of Mathematics,\\
PO Box 80203, Jeddah, 21589, Saudi Arabia.}\\
E-mail: atali71@yahoo.com\\
\normalsize{$^\$$ Mathematics Department,\\ Faculty of Science, Al-Azhar University,\\
Nasr city, 11884, Cairo, Egypt}\\
\vspace{2mm}
\normalsize{$^\ddag$ Department of Physics, Anand Engineering College,\\
Keetham, Agra - 282 007, India.\\
E-mail: abanilyadav@yahoo.co.in}\\
\vspace{2mm}
\normalsize{$^\S$ Mathematics Department,\\
Faculty of Science, Sohag University, Egypt.}

\end{center}
\begin{abstract} The present study deals with the inhomogeneous plane symmetric models in
scalar - tensor theory of gravitation. We used symmetry group analysis method to solve the
field equations analytically. A new class of similarity solutions have been obtained
by considering the inhomogeneous nature of metric potential. The physical behavior and
geometrical aspects of the derived models are also discussed.
\end{abstract}

\section{Introduction }
In recent years, modifications of general relativity are attracting more attention 
to explain the late time cosmic acceleration of universe. This late time cosmic 
accelerated expansion of universe has been confirmed by high red-shift supernovae experiments 
(Riess et al 1998; Perlmutter et al 1999; Bennet et al 2003). Broadly, the model building 
undertaken in the literature to capture the alternative theory of gravitation can be 
classified into two categories: dimensional scalar field and non dimensional scalar field model. 
In 1961, Brans and Dicke formulated the scalar-tensor theories of gravitation on the basis 
of couping between an adequate tensor field and scalar field $\phi$.  The scalar field has a
dimension of $G^{-1}$ where $G$ is the gravitational constant. Therefore $\phi^{-1}$ play the role of $G(t)$.
This theory successfully describes the Mach's principle but fails to explain the missing matter problems 
and absolute properties of space.  Later on Saez and Ballester (1985) developed a scalar-tensor theory
in which the metric is coupled with a dimensionless scalar
field in a simple manner. This coupling gives a satisfactory
description of weak fields. The SB theory of gravitation solves missing matter problem in non flat FRW cosmologies 
and removes the graceful exist problem in inflation era. In the literature, Singh and Agarwal (1991), 
Reddy et al (2006), Socorro et al (2010), Jamil et al (2012) and recently Yadav (2013) have studied the 
some aspects of SB theory of gravitation in different physical contexts.\\

The recent observations suggest that the matter distribution in the present universe is on 
the whole isotropic and homogeneous. But on the theoretical ground, the universe could have not 
had such smoothed out picture. Close to big bang singularity, the assumption of spherically symmetric 
and isotropy can not be strictly valid. Therefore inhomogeneous cosmological models play an important 
role to study the essential features of universe such as process of homogenization and 
formation of galaxies at early stage of evolution. So in literature, many authors consider plane symmetry, which is
less restrictive than spherical symmetry and provides an avenue to study inhomogeneities in early universe. 
Rendall (1995), Da Silva and Wang (1998),
Anguige (2000), Nouri-Zonoz and Tavanfar (2001), Pradhan et al. (2003, 2007) and Yadav (2011) 
have studied the plane symmetric and inhomogeneous cosmological models in different physical context. 
In 2008, Marra and Paakonen (2008) and recently Ali and Yadav (2013) have presented 
the exact solution which governs the dynamics of inhomogeneous universe.\\

The non linear equations are widely used as a model to describes the complex physical phenomenon in general 
relativity and fluid mechanics. According to Sabbagh and Ali (2008) the exact solution of non 
linear partial differential equations is important for the study of non linear physical phenomenon. 
The symmetry groups are defined as the groups of continuous
transformations that leave a given family of equations invariant (2009, 2013). 
In this paper, we apply the symmetry group analysis method for a particular problem in 
the inhomogeneous plane symmetric models in a scalar tensor theory. The similarity solutions are 
quite popular because they result in the reduction of the independent variables of the problem. 
In our case, the problem under investigation is the system of second order nonlinear PDEs. Hence, 
any similarity solution will transform the system of nonlinear PDEs into a system of ODEs.\\

In our paper, we find a new class of exact solution for inhomogeneous universe in scalar-tensor
theory of gravitation. The paper is organized as follows: In section 2, we have provided the
metric and field equation in connection to the proposed model for inhomogeneous universe. Section
3 and 4 are delt respectively, with the symmetry group analysis method and similarity solution
for the models under consideration. Some concluding remarks are made in section 5.

\section{The metric and field equations}

We consider the general plane symmetric metric in the form
\begin{equation}  \label{u21}
ds^2=E\,\big(dz^2-dt^2\big)+G\,\big(dx^2+dy^2\big),
\end{equation}
where $E$ and $G$ are functions of $z$ and $t$. The field equations in Saez and Ballester (1985) theory are
\begin{equation}  \label{u22}
R_{ij}-\dfrac{1}{2}\,g_{ij}\,R-\omega\,\phi^n\,\Big(\phi_{,i}\,\phi_{,j}-\dfrac{1}{2}\,g_{ij}\,\phi_{,k}\,\phi^{,k}\Big)=K\,T_{ij},
\end{equation}
where $\phi$ is a scalar function of $z$ and $t$ while $n$ is an arbitrary exponent constant and $\omega$ is dimensionless coupling constant. The scalar field $\phi$ satisfies the equation
\begin{equation}  \label{u23}
2\,\phi^n\,\phi^{;i}_{;i}+n\,\phi^{n-1}\,\phi_{,k}\,\phi^{,k}=0.
\end{equation}
In the case of a perfect fluid distribution, the energy-momentum tensor $T_{ij}$ is given by
\begin{equation}  \label{u24}
T_{ij}=(p+\rho)\,v_i\,v_j+p\,g_{ij},
\end{equation}
where $\rho$ is the matter energy density, $p$ the pressure and $v^i$ the fluid four velocity vector. As a consequence of Bianchi identities, the equations of the motion are
\begin{equation}  \label{u25}
T^{ij}_{;j}\,=\,0.
\end{equation}
For the metric (\ref{u21}), the field equations (\ref{u22}), (\ref{u23}) and (\ref{u25}) in co-moving coordinates leads to

\begin{equation}  \label{u26}
\begin{array}{ll}
F_1\,=\,\dfrac{G_{zt}}{G}-\dfrac{G_z\,G_t}{2\,G^2}-\dfrac{E_z\,G_t+E_t\,G_z}{2\,E\,G}+\omega\,\phi^{n}\,\phi_z\,\phi_t\,=\,0,
  \end{array}
\end{equation}

\begin{equation}  \label{u27}
\begin{array}{ll}
F_2\,=\,\dfrac{G_{zz}+G_{tt}}{G}+\dfrac{E_{zz}-E_{tt}}{E}-\dfrac{G_z^2}{G^2}-\dfrac{E_z\,G_t+E_t\,G_z}{E\,G}+\dfrac{E_t^2-E_z^2}{E^2}
+2\,\omega\,\phi^{n}\,\phi_z^2\,=\,0,
  \end{array}
\end{equation}

\begin{equation}  \label{u28}
\begin{array}{ll}
F_3\,=\,\dfrac{\phi_{zz}-\phi_{tt}}{\phi}+\dfrac{n}{2}\Big(\dfrac{\phi_z^2-\phi_t^2}{\phi^2}\Big)+\dfrac{G_z\,\phi_z-G_t\,\phi_t}{G\,\phi}\,=\,0,
  \end{array}
\end{equation}

\begin{equation}  \label{u29}
\begin{array}{ll}
        K\,E\,p+\dfrac{G_{tt}}{G}-\dfrac{G_z^2+G_t^2}{4\,G^2}-\dfrac{E_z\,G_t+E_t\,G_z}{E\,G}+\dfrac{1}{2}\,\omega\,\phi^n\,(\phi_z^2+\phi_t^2)\,=\,0,
  \end{array}
\end{equation}

\begin{equation}  \label{u210}
\begin{array}{ll}
        K\,E\,\rho+\dfrac{G_{zz}}{G}-\dfrac{G_z^2+G_t^2}{4\,G^2}-\dfrac{E_z\,G_t+E_t\,G_z}{E\,G}+\dfrac{1}{2}\,\omega\,\phi^n\,(\phi_z^2+\phi_t^2)\,=\,0,
  \end{array}
\end{equation}

\begin{equation}  \label{u211}
\begin{array}{ll}
p_z+(\rho+p)\Big(\dfrac{E_z}{2\,E}\Big)\,=\,0,
  \end{array}
\end{equation}

\begin{equation}  \label{u212}
\begin{array}{ll}
\rho_t+(\rho+p)\Big(\dfrac{E_t}{2\,E}+\dfrac{G_t}{G}\Big)\,=\,0,
  \end{array}
\end{equation}

\section{Symmetry analysis method}

Equations (\ref{u26})-(\ref{u212}) are highly non-linear partial differential equations and
hence is very difficult to solve them, as there exist no standard method for their solution.
The system (\ref{u26})-(\ref{u28}) are nonlinear partial differential equations of second order
for the three unknowns $E$, $G$ and $\phi$. If we solve this system, then we can get the solution of
the field equations. In order to obtain an exact solutions of system of nonlinear partial differential
equations (\ref{u26})-(\ref{u28}), we will use the symmetry
analysis method. For this we write
\begin{equation}\label{u31}
\left\{
\begin{array}{ll}
x_i^{*}=x_i+\epsilon\,\xi_{i}(x_j,u_{\beta})+\bold{o}(\epsilon^2),\\
u_{\alpha}^{*}=u_{\alpha}+\epsilon\,\eta_{\alpha}(x_j,u_{\beta})+\bold{o}(\epsilon^2),
\end{array}
\right.
\,\,\,i,j=1,2,\,\,\,\alpha,\beta=1,2,3,
\end{equation}
as the infinitesimal Lie point transformations. We have assumed
that the system (\ref{u26})-(\ref{u28}) is invariant under the transformations given in
Eq. (\ref{u31}). The corresponding infinitesimal generator of Lie groups
(symmetries) is given by
\begin{equation}\label{u32}
 X=\sum_{i=1}^{2}\xi_{i}\dfrac{\partial}{\partial x_{i}}+\sum_{\alpha=1}^{3}\eta_{\alpha}
 \dfrac{\partial}{\partial u_{\alpha}},
 \end{equation}
where $x_1=z$, $x_2=t$, $u_1=E$, $u_2=G$ and $u_3=\phi$. The coefficients $\xi_{1}$, $\xi_{2}$, $\eta_{1}$, $\eta_2$ and $\eta_{3}$ are the functions of $z$, $t$, $E$, $G$ and $\phi$.
These coefficients are the components of infinitesimals symmetries
corresponding to $z$, $t$, $E$, $G$ and $\phi$ respectively, to be determined from the invariance conditions:
\begin{equation}\label{u33}
{\text{Pr}}^{(2)}\,X\Big(F_m\Big)|_{F_m=0}=0,
\end{equation}
where $F_m=0,\,m=1,2,3$ are the system (\ref{u26})-(\ref{u28}) under study and
${\text{Pr}}^{(2)}$ is the second prolongation of the symmetries $X$.
Since our equations (\ref{u26})-(\ref{u28}) are at most of order two, therefore, we
need second order prolongation of the infinitesimal generator
in Eq. (\ref{u33}). It is worth noting that, the $n$-th order prolongation is given by:
\begin{equation}\label{u34}
{\text{Pr}}^{(n)}\,X=X+\sum_{s=1}^{n}\,\sum_{\alpha=1}^{3}\,\eta_{\alpha,i_1i_2...i_s}\,\dfrac{\partial}{\partial u_{\alpha,i_1i_2...i_s}},
\end{equation}
where
\begin{equation}\label{u35}
\eta_{\alpha,i_1i_2...i_s}=D_{i_1i_2...i_s}\Big[\eta_{\alpha}-\sum_{i=1}^{2}
\,\xi_i\,u_{\alpha,i}\Big]+\sum_{i=1}^{2}\,\xi_{i}\,u_{\alpha,i_1i_2...i_si}\,.
\end{equation}
The operator $D_{i_1i_2...i_s}$ is called the {\it total derivative} ({\it Hach operator}) and taken the following
form:
\begin{equation}\label{u36}
D_i=\dfrac{\partial}{\partial x_i}+\sum_{s=1}^{n}\,\sum_{\alpha=1}^{3}\,u_{\alpha,i_1i_2...i_s}\,
\dfrac{\partial}{\partial u_{\alpha,i_1i_2...i_s}},
\end{equation}
where $D_{ij}=D_{ji}$ and $u_{\alpha,i}=\frac{\partial u_{\alpha}}{\partial x_{i}}$.

Expanding Eqs. (\ref{u33}) with the original system of Eqs. (\ref{u26})-(\ref{u28}) to eliminate $E_{zz}$, $G_{zt}$
and $\phi_{zz}$ while we set the coefficients involving $E_{z}$, $E_{t}$, $E_{zt}$, $E_{tt}$, $G_{z}$,
$G_{t}$, $G_{zz}$, $G_{tt}$,  $\phi_{z}$, $\phi_{t}$, $\phi_{zt}$, $\phi_{tt}$ and
various products to zero give rise the essential set of over-determined
equations. Solving the set of these determining equations, the components of symmetries takes the following form:
\begin{equation}\label{u37}
\xi_{1}=c_1\,z+c_2,\,\,\,\xi_{2}=c_1\,t+c_3,\,\,\,\eta_{1}=c_4\,E ,\,\,\,\eta_{2}=c_5\,G,\,\,\,\eta_3=c_6\,\phi^{-n/2},
\end{equation}
where $c_i,\,i=1,2,...,6$ are an arbitrary constants.

\section{Similarity solutions}

The characteristic equations corresponding to the symmetries (\ref{u37}) are given by:
\begin{equation}\label{u41}
\dfrac{dz}{c_1\,z+c_2}=\dfrac{dt}{c_1\,t+c_3}=\dfrac{dE}{c_4\,E}=\dfrac{dG}{c_5\,G}=\dfrac{d\phi}{c_6\,\phi^{-n/2}}.
\end{equation}
By solving the above system, we have the following four cases:\\

\textbf{Case (1):} When $n\,=\,-2$ and $c_1\,=\,0$, the similarity variable and similarity 
functions can be written as the following:
\begin{equation}\label{u42-1}
\left\{
\begin{array}{ll}
\xi=a\,z+b\,t,\\
E(z,t)=\Psi(\xi)\,\exp[c\,z],\\
G(z,t)=\Omega(\xi)\,\exp[d\,z],\\
\phi(z,t)=\Phi(\xi)\,\exp[k\,z],
\end{array}
\right.
\end{equation}
where $a=c_3$, $b=-c_2$, $c=\dfrac{c_4}{c_2}$, $d=\dfrac{c_5}{c_2}$ and $k=\dfrac{c_6}{c_2}$ are an 
arbitrary constants. Substituting the transformations (\ref{u42-1}) in the field Eqs. (\ref{u26})-(\ref{u28}) lead
to the following system of ordinary differential equations:

\begin{equation}\label{u43-1}
\begin{array}{ll}
\Big(\dfrac{b^2-a^2}{a^2}\Big)\,\Big(\dfrac{\Psi\,\Psi''-\Psi'^2}{\Psi^2}\Big)
-\Big(\dfrac{b^2+a^2}{a^2}\Big)\,\Big(\dfrac{\Psi\,\Omega''-\Psi'\,\Omega'}{\Psi\,\Omega}\Big)\\
\\
\,\,\,\,\,\,\,\,\,\,\,\,\,\,\,\,\,\,\,\,\,\,\,\,\,\,\,\,\,\,\,\,\,\,\,
+\dfrac{\Omega'^2}{\Omega^2}+\dfrac{d\,\Psi'}{a\,\Psi}+\dfrac{c\,\Omega'}{a\,\Omega}=2\,\omega\,\Big(\dfrac{\Phi'}{\Phi}+\dfrac{k}{a}\Big)^2
-\dfrac{c\,d}{a^2},
\end{array}
\end{equation}

\begin{equation}\label{u44-1}
\begin{array}{ll}
\dfrac{\Omega''}{\Omega}-\dfrac{\Omega'^2}{2\,\Omega^2}-\dfrac{\Psi'\,\Omega'}{\Psi\,\Omega}+\dfrac{\omega\,\Phi'^2}{\Phi^2}=\dfrac{d\,\Psi'}{2\,a\,\Psi}
+\dfrac{(c-d)\,\Omega'}{2\,a\,\Omega}-\dfrac{k\,\omega\,\Phi'}{a\,\Phi},
\end{array}
\end{equation}

\begin{equation}\label{u45-1}
\begin{array}{ll}
\Big(\dfrac{b^2-a^2}{a}\Big)\,\Big[\dfrac{\Phi''}{\Phi}+\dfrac{\Omega'\,\Phi'}{\Omega\,\Phi}-\dfrac{\Phi'^2}{\Phi^2}\Big]
=\dfrac{d\,\Phi'}{\Phi}+\dfrac{k\,\Omega'}{\Omega}+\dfrac{d\,k}{a}.
\end{array}
\end{equation}

The equations (\ref{u43-1})-(\ref{u45-1}) are non-linear ordinary differential 
equations which is very difficult to solve. However, it is worth noting that, this equations 
can be solved when $b=a$. In this case, the equation (\ref{u45-1}) takes the form:
\begin{equation}\label{u46-1}
\begin{array}{ll}
\dfrac{d\,\Phi'}{\Phi}+\dfrac{k\,\Omega'}{\Omega}+\dfrac{d\,k}{a}\,=\,0.
\end{array}
\end{equation}
By integration the above equation, we get:
\begin{equation}\label{u47-1}
\begin{array}{ll}
\Phi(\xi)=q_3\,\Omega^{-k/d}(\xi)\,\exp\Big[-\Big(\dfrac{k}{a}\Big)\,\xi\Big],
\end{array}
\end{equation}
where $q_3$ is an arbitrary constant of integration. Under the solution (\ref{u47-1}), 
the equation (\ref{u44-1}) reduce to the following ordinary differential equation
\begin{equation}\label{u48-1}
\dfrac{\Psi'}{\Psi}=\dfrac{a\,(2\,k^2\,\omega-d^2)\,\Omega'^2+d\,\Omega\,\big[
2\,a\,d\,\Omega''+(2\,k^2\,\omega+d^2-c\,d)\,\Omega'\big]}{d^2\,\Omega\,\big[2\,a\,\Omega'+d\,\Omega\big]}.
\end{equation}

Integrate the above equation for the function $\Psi$, we have:

\begin{equation}\label{u49-1}
\Psi(\xi)=q_1\,\exp\Bigg[\int\dfrac{a\,(2\,k^2\,\omega-d^2)\,\Omega'^2+d\,\Omega\,\big[
2\,a\,d\,\Omega''+(2\,k^2\,\omega+d^2-c\,d)\,\Omega'\big]}{d^2\,\Omega\,\big[2\,a\,\Omega'+d\,\Omega\big]}\,d\xi\Bigg],
\end{equation}
where $q_1$ is an arbitrary constant of integration. From (\ref{u47-1}) and (\ref{u49-1}), 
the equation (\ref{u43-1}) becomes

\begin{equation}\label{u410-1}
\begin{array}{ll}
a\,(2\,k^2\,\omega+d^2)\,\Omega'+c\,d^2\,\Omega=0.
\end{array}
\end{equation}
The general solution of the equation (\ref{u410-1}) is:
\begin{equation}\label{u411-1}
\begin{array}{ll}
\Omega(\xi)=q_2\,\exp\Big[-\Big(\dfrac{c\,d^2}{a\,(2\,k^2\,\omega+d^2)}\Big)\,\xi\Big].
\end{array}
\end{equation}
where $q_2$ is an arbitrary constant of integration. Using (\ref{u411-1}), (\ref{u49-1}), (\ref{u47-1}) 
and the transformation (\ref{u42-1}), we have the following solution of the field equation
\begin{equation}\label{u412-1}
\left\{
\begin{array}{ll}
E(z,t)=q_1\,\exp\big[-c\,t\big],\\
\\
G(z,t)=q_2\,\exp\Big[d\Big(z-\dfrac{c\,d\,(t+z)}{2\,k^2\,\omega+d^2}\Big)\Big],\\
\\
\phi(z,t)=\tilde{q}_3\,\exp\Big[k\Big(\dfrac{c\,d\,(t+z)}{2\,k^2\,\omega+d^2}-t\Big)\Big],
\end{array}
\right.
\end{equation}
where $\tilde{q}_3=q_3\,q_2^{-k/d}$. The metric of the corresponding solution can be written in the following form:
\begin{equation}  \label{s1}
\begin{array}{ll}
ds_{1}^2=q_1\,\exp\big[-c\,t\big]\,\big(dz^2-dt^2\big)+q_2\,\exp\Big[d\Big(z-\dfrac{c\,d\,(t+z)}{2\,k^2\,\omega+d^2}\Big)\Big]\,\big(dx^2+dy^2\big),
\end{array}
\end{equation}
where $q_1$, $q_2$, $c$, $d$, $k$ and $\omega$ are an arbitrary constants.\\

The energy density and pressure for the model (\ref{s1}) are given by
\begin{equation}  \label{u413-1}
\begin{array}{ll}
\rho(z,t)=\dfrac{(2\,k^2\,\omega+3\,d^2)\,(2\,c\,d-d^2-2\,k^2\,\omega)\,\exp[c\,t]}{4\,q_1\,K\,(2\,k^2\,\omega+d^2)},
\end{array}
\end{equation}

\begin{equation}  \label{u414-1}
\begin{array}{ll}
p(z,t)=\dfrac{(2\,k^2\,\omega-d^2)\,(2\,c\,d-d^2-2\,k^2\,\omega)\,\exp[c\,t]}{4\,q_1\,K\,(2\,k^2\,\omega+d^2)}.
\end{array}
\end{equation}
The spatial volume $V$ is given by
\begin{equation}  \label{u415-1}
\begin{array}{ll}
V(z,t)=q_1\,q_2\,\exp\Big[d\,z-c\Big(t+\dfrac{d^2\,(z+t)}{2\,k^2\,\omega+d^2}\Big)\Big].
\end{array}
\end{equation}
The scalar expansion $\Theta$ and the shear scalar $\sigma$ are given by Collins and Wainwright \cite{collin1}:
\begin{equation}  \label{u416-1}
\begin{array}{ll}
\Theta=-\Big(\dfrac{c\,(2\,k^2\,\omega+3\,d^2)}{2\,\sqrt{q_1}\,(2\,k^2\,\omega+d^2)}\Big)\,\exp\Big[\dfrac{c\,t}{2}\Big],
\end{array}
\end{equation}

\begin{equation}  \label{u417-1}
\begin{array}{ll}
\sigma=\dfrac{c^2\,k^4\,\omega^2\,\exp[c\,t]}{3\,q_1\,(2\,k^2\,\omega+d^2)}.
\end{array}
\end{equation}
The deceleration parameter $\mathbf{q}$ is given by:
\begin{equation}  \label{u418-1}
\begin{array}{ll}
\mathbf{q}=\dfrac{c^4\,k^2\,\omega\,(2\,k^2\,\omega+3\,d^2)^3\,\exp[2\,c\,t]}{4\,q_1^2\,(2\,k^2\,\omega+d^2)^4}.
\end{array}
\end{equation}

\textbf{Case (2):} When $n\,\neq\,-2$ and $c_1\,=\,0$, the similarity variable and similarity functions can be written as the following:
\begin{equation}\label{u42-2}
\left\{
\begin{array}{ll}
\xi=a\,z+b\,t,\\
E(z,t)=\Psi(\xi)\,\exp[c\,z],\\
G(z,t)=\Omega(\xi)\,\exp[d\,z],\\
\phi(z,t)=\Big(\Phi(\xi)+\tilde{k}\,z\Big)^{\dfrac{2}{n+2}},
\end{array}
\right.
\end{equation}
where $a=c_3$, $b=-c_2$, $c=\dfrac{c_4}{c_2}$, $d=\dfrac{c_5}{c_2}$ and $\tilde{k}=\dfrac{(n+2)\,c_6}{2\,c_2}$ are an arbitrary constants. Substituting the transformations (\ref{u42-2}) in the field Eqs. (\ref{u26})-(\ref{u28}) lead
to the following system of ordinary differential equations:

\begin{equation}\label{u43-2}
\begin{array}{ll}
\Big(\dfrac{b^2-a^2}{a^2}\Big)\,\Big(\dfrac{\Psi\,\Psi''-\Psi'^2}{\Psi^2}\Big)
-\Big(\dfrac{b^2+a^2}{a^2}\Big)\,\Big(\dfrac{\Psi\,\Omega''-\Psi'\,\Omega'}{\Psi\,\Omega}\Big)\\
\\
\,\,\,\,\,\,\,\,\,\,\,\,\,\,\,\,\,\,\,\,\,\,\,\,\,\,\,\,\,\,\,\,\,\,\,
+\dfrac{\Omega'^2}{\Omega^2}+\dfrac{d\,\Psi'}{a\,\Psi}+\dfrac{c\,\Omega'}{a\,\Omega}=\dfrac{8\,\omega}{a^2\,(n+2)^2}\,\Big(a\,\Phi'+\tilde{k}\Big)^2
-\dfrac{c\,d}{a^2},
\end{array}
\end{equation}

\begin{equation}\label{u44-2}
\begin{array}{ll}
\dfrac{\Omega''}{\Omega}-\dfrac{\Omega'^2}{2\,\Omega^2}-\dfrac{\Psi'\,\Omega'}{\Psi\,\Omega}+\dfrac{4\,\omega\,\Phi'^2}{(n+2)^2}=\dfrac{d\,\Psi'}{2\,a\,\Psi}
+\dfrac{(c-d)\,\Omega'}{2\,a\,\Omega}-\dfrac{4\,\tilde{k}\,\omega\,\Phi'}{a\,(n+2)^2},
\end{array}
\end{equation}

\begin{equation}\label{u45-2}
\begin{array}{ll}
\Big(\dfrac{b^2-a^2}{a}\Big)\,\Big[\dfrac{\Phi''}{\Phi}+\dfrac{\Omega'\,\Phi'}{\Omega\,\Phi}\Big]
=\dfrac{1}{\Phi}\,\Big(d\,\Phi'+\dfrac{\tilde{k}\,\Omega'}{\Omega}+\dfrac{d\,k}{a}\Big).
\end{array}
\end{equation}

The equations (\ref{u43-2})-(\ref{u45-2}) are non-linear ordinary differential equations which is very difficult to solve.
However, it is worth noting that, this equations can be solved when $b=a$. In this case, the equation (\ref{u45-2})
takes the form:
\begin{equation}\label{u46-2}
\begin{array}{ll}
d\,\Phi'+\dfrac{\tilde{k}\,\Omega'}{\Omega}+\dfrac{d\,\tilde{k}}{a}\,=\,0.
\end{array}
\end{equation}
By integration the above equation, we get:
\begin{equation}\label{u47-2}
\begin{array}{ll}
\Phi(\xi)=q_3-\Big(\dfrac{\tilde{k}}{a}\Big)\,\xi-\dfrac{\tilde{k}}{d}\,\ln\Big[\Omega(\xi)\Big],
\end{array}
\end{equation}
where $q_3$ is an arbitrary constant of integration. Under the solution (\ref{u47-2}), the equation
(\ref{u44-2}) reduce to the following ordinary differential equation
\begin{equation}\label{u48-2}
\dfrac{\Psi'}{\Psi}=\dfrac{a\,\Big[8\,\tilde{k}^2\,\omega-d^2\,(n+2)^2\Big]\,\Omega'^2+d\,\Omega\,\Big[
2\,a\,d\,\Omega''+\big[8\,\tilde{k}^2\,\omega-d\,(c-d)\,(n+2)^2\big]\,\Omega'\Big]}{d^2\,(n+2)^2\,\Omega\,\big[2\,a\,\Omega'+d\,\Omega\big]}.
\end{equation}

Integrate the above equation for the function $\Psi$, we have:

\begin{equation}\label{u49-2}
\begin{array}{ll}
\Psi(\xi)=q_1\,\exp\Bigg[\int\Bigg(\dfrac{a\,\Big[8\,\tilde{k}^2\,\omega-d^2\,(n+2)^2\Big]\,\Omega'^2}{d^2\,(n+2)^2\,\Omega\,\big[2\,a\,\Omega'+d\,\Omega\big]}\\
\\
\,\,\,\,\,\,\,\,\,\,\,\,\,\,\,\,\,\,\,\,\,\,\,\,\,\,\,\,\,\,\,\,\,\,\,\,\,\,\,\,
\,\,\,\,\,\,\,\,\,\,\,\,\,\,\,\,\,\,\,\,
+\dfrac{d\,\Omega\,\Big[
2\,a\,d\,\Omega''+\big[8\,\tilde{k}^2\,\omega-d\,(c-d)\,(n+2)^2\big]\,\Omega'\Big]}{d^2\,(n+2)^2\,\Omega\,\big[2\,a\,\Omega'+d\,\Omega\big]}\Bigg)\,d\xi\Bigg],
\end{array}
\end{equation}
where $q_1$ is an arbitrary constant of integration. From (\ref{u47-2}) and (\ref{u49-2}),
the equation (\ref{u43-2}) becomes

\begin{equation}\label{u410-2}
\begin{array}{ll}
a\,\Big[8\,\tilde{k}^2\,\omega+d^2\,(n+2)^2\Big]\,\Omega'+c\,d^2\,(n+2)^2\,\Omega=0.
\end{array}
\end{equation}
The general solution of the equation (\ref{u410-2}) is:
\begin{equation}\label{u411-2}
\begin{array}{ll}
\Omega(\xi)=q_2\,\exp\Bigg[-\Bigg(\dfrac{c\,d^2\,(n+2)^2}{a\,\Big[8\,\tilde{k}^2\,\omega+d^2\,(n+2)^2\Big]}\Bigg)\,\xi\Bigg].
\end{array}
\end{equation}
where $q_2$ is an arbitrary constant of integration. Using (\ref{u411-2}), (\ref{u49-2}), (\ref{u47-2}) and
the transformation (\ref{u42-2}), we have the following solution of the field equation
\begin{equation}\label{u412-2}
\left\{
\begin{array}{ll}
E(z,t)=q_1\,\exp\big[-c\,t\big],\\
\\
G(z,t)=q_2\,\exp\Bigg[d\Bigg(z-\dfrac{c\,d\,(n+2)^2\,(t+z)}{8\,\tilde{k}^2\,\omega+d^2\,(n+2)^2}\Bigg)\Bigg],\\
\\
\phi(z,t)=\Bigg[\tilde{q}_3-\tilde{k}\,\Bigg(t-\dfrac{c\,d\,(n+2)^2\,(t+z)}{8\,\tilde{k}^2\,\omega+d^2\,(n+2)^2}\Bigg)\Bigg]^{\dfrac{2}{n+2}},
\end{array}
\right.
\end{equation}
where $\tilde{q}_3=q_3-\dfrac{k}{d}\ln[q_2]$. The metric of the corresponding solution can be written in the following form:
\begin{equation}  \label{s2}
\begin{array}{ll}
ds_{2}^2=q_1\,\exp\big[-c\,t\big]\,\big(dz^2-dt^2\big)
+q_2\,\exp\Bigg[d\Bigg(z-\dfrac{c\,d\,(n+2)^2\,(t+z)}{8\,\tilde{k}^2\,\omega+d^2\,(n+2)^2}\Bigg)\Bigg]\,\big(dx^2+dy^2\big),
\end{array}
\end{equation}
where $q_1$, $q_2$, $c$, $d$, $\tilde{k}$, $n$ and $\omega$ are an arbitrary constants.\\

\textbf{Remark (1):} The metric (\ref{s2}) is equal the metric (\ref{s1}) when $\tilde{k}=\dfrac{(n+2)\,k}{2}$. Then the solution in the case 1 and case 2 are the same such that:

\begin{equation}\label{u413-2}
\left\{
\begin{array}{ll}
\phi(z,t)=\tilde{q}_3\,\exp\Big[k\Big(\dfrac{c\,d\,(t+z)}{2\,k^2\,\omega+d^2}-t\Big)\Big],\,\,\,\mathrm{When}\,\,\,n\,=\,-2,\\
\\
\phi^{\frac{n}{2}+1}(z,t)=\tilde{q}_3-\dfrac{(n+2)\,k}{2}\,\Bigg(t-\dfrac{c\,d\,(t+z)}{2\,k^2\,\omega+d^2}\Bigg),
\,\,\,\mathrm{When}\,\,\,n\,\neq\,-2.
\end{array}
\right.
\end{equation}

\textbf{Case (3):} When $n\,=\,-2$ and $c_1\,\neq\,0$, the similarity variable and similarity functions can be written as the following:
\begin{equation}\label{u42-3}
\left\{
\begin{array}{ll}
\xi\,=\,\dfrac{z+a}{t+b},\\
E(z,t)\,=\,(z+a)^c\,\Psi(\xi),\\
G(z,t)\,=\,(z+a)^d\,\Omega(\xi),\\
\phi(z,t)\,=\,(z+a)^k\,\Phi(\xi),
\end{array}
\right.
\end{equation}
where $a=\dfrac{c_2}{c_1}$, $b=\dfrac{c_3}{c_1}$, $c=\dfrac{c_4}{c_1}$, $d=\dfrac{c_5}{c_1}$ and $k=\dfrac{c_6}{c_1}$ are an arbitrary constants. Substituting the transformations (\ref{u42-3}) in the field Eqs. (\ref{u26})-(\ref{u28}) lead
to the following system of ordinary differential equations:

\begin{equation}\label{u43-3}
\begin{array}{ll}
\xi^2\,(\xi^2-1)\,\Big(\dfrac{\Psi\,\Psi''-\Psi'^2}{\Psi^2}\Big)
-\xi^2\,(\xi^2+1)\,\Big(\dfrac{\Psi\,\Omega''-\Psi'\,\Omega'}{\Psi\,\Omega}\Big)+\xi^2\,\Big(\dfrac{\Omega'^2}{\Omega^2}\Big)\\
\\
\,\,\,\,\,\,\,\,\,\,\,\,\,\,\,\,\,\,\,\,\,\,\,\,\,
+\xi\,(d+2\,\xi^2)\,\dfrac{\Psi'}{\Psi}+\xi\,(c-2\,\xi^2)\,\dfrac{\Omega'}{\Omega}
=2\,\omega\,\xi^2\,\Big(\dfrac{\Phi'}{\Phi}+\dfrac{k}{\xi}\Big)^2-c\,d-d-c,
\end{array}
\end{equation}

\begin{equation}\label{u44-3}
\begin{array}{ll}
\xi\,\Big[\dfrac{\Omega''}{\Omega}-\dfrac{\Omega'^2}{2\,\Omega^2}-\dfrac{\Psi'\,\Omega'}{\Psi\,\Omega}+\dfrac{\omega\,\Phi'^2}{\Phi^2}\Big]
=\dfrac{d\,\Psi'}{2\,\Psi}
+\dfrac{(c-d-2)\,\Omega'}{2\,\Omega}-\dfrac{k\,\omega\,\Phi'}{\Phi},
\end{array}
\end{equation}

\begin{equation}\label{u45-3}
\begin{array}{ll}
\xi^2\,(1-\xi^2)\,\Big[\dfrac{\Phi''}{\Phi}+\dfrac{\Omega'\,\Phi'}{\Omega\,\Phi}-\dfrac{\Phi'^2}{\Phi^2}\Big]
=\xi\,(2\,\xi^2-d)\,\Big(\dfrac{\Phi'}{\Phi}\Big)-k\,\xi\,\Big(\dfrac{\Omega'}{\Omega}\Big)-k\,(d-1).
\end{array}
\end{equation}

If one solves the system of second order non-linear ordinary differential
equations (\ref{u43-3})-(\ref{u45-3}), he can obtain the exact solutions of the original
field equations (\ref{u26})-(\ref{u28}) corresponding to reduction (\ref{u42-3}).
In general, one can not solve the system of equations (\ref{u43-3})-(\ref{u45-3}).
So, in order to solve the problem completely, we have to choose some special cases as following:\\

We assume the solution of the function $\Phi(\xi)$ in the form:
\begin{equation}\label{u46-3}
\begin{array}{ll}
\Phi(\xi)=q_3\,\xi^{q_4},
\end{array}
\end{equation}
where $q_3$ and $q_4$ are arbitrary non-zero constants.\\

\textbf{Case (3.1):}  When $d\,\neq\,2$\\
Now solving equation (\ref{u46-3}) with equations (\ref{u44-3}) and (\ref{u45-3}), we obtain
\begin{equation}\label{u47-3}
\begin{array}{ll}
\Omega(\xi)=q_2\,\xi^{1-d}\,\Big[q_4\,\xi^2-(q_4+k)\Big]^{\dfrac{d-2}{2}},
\end{array}
\end{equation}

\begin{equation}\label{u48-3}
\begin{array}{ll}
\Psi(\xi)=q_1\,\xi^{\alpha_1}\,\Big[(q_4+k)-q_4\,\xi^2\Big]^{\dfrac{d-6}{4}}\,\Big[(q_4+k)+q_4\,\xi^2\Big]^{\alpha_2},
\end{array}
\end{equation}
where $q_1$ and $q_2$ are arbitrary non-zero constants while $\alpha_1=\dfrac{(d-1)\,(1-c)-2\,q_4\,\omega\,(q_4+k)}{d-2}$ and \\
$\alpha_2=\dfrac{d\,(d-2\,c)-4\,(d-2)-8\,q_4\,\omega\,(q_4+k)}{4\,(2-d)}$.\\ 
Therefore, the equation (\ref{u43-3}) can be written in the form
\begin{equation}\label{u49-3}
\begin{array}{ll}
Q_0+Q_1\,\xi^2+Q_2\,\xi^4+Q_3\,\xi^6+Q_4\,\xi^8\,=\,0,
\end{array}
\end{equation}
where
\begin{equation}\label{u410-3}
\left\{
\begin{array}{ll}
Q_0=(q_4+k)^4\,\Big[d\,\big[3-2\,\omega\,(q_4+k)^2\big]-4\,\big[1-k\,\omega\,(q_4+k)\big]-2\,c\Big],\\
Q_1=(d-2)\,(q_4+k)^3\,\Big[(q_4+k)\,\big[2\,q_4\,\omega\,(2\,q_4+k)+d\,(c+2)\big]\\
\,\,\,\,\,\,\,\,\,\,\,\,\,\,\,\,\,\,\,\,\,\,\,\,\,\,\,\,\,\,\,\,\,\,\,\,\,\,\,\,\,\,\,\,\,\,\,\,\,\,\,\,\,\,
\,\,\,\,\,\,\,\,\,\,\,\,\,\,\,\,\,\,\,\,\,\,\,\,\,\,\,\,\,\,\,\,\,\,\,\,\,\,\,\,\,\,\,\,\,\,\,\,\,\,\,\,\,\,
-k\,(1+c+d^2)-2\,q_4\Big],\\
Q_2=q_4\,(q_4+k)^2\,\Big[1\,(d+2)\,k^2\,q_4\,\omega+k\,\big[
2\,d\,q_4^2\,\omega+28\,q_4^2\,\omega+c\,(d+2)\\
\,\,\,\,\,\,\,\,\,\,
-d^2\,(d-2)+9\,d-22\big]+q_4\,\big[
c\,(4+6\,d-d^2)-4\,(5-3\,d+d^2-6\,q_4^2\,\omega)\big]\Big],\\
Q_3=-q_4^2\,(q_4+k)\,\Big[2\,(6+d)\,k^2\,q_4\,\omega+k\,\big[
2-d-2\,d^2+c\,(2+d+d^2)\\
\,\,\,\,\,\,\,\,\,\,
+36\,q_4^2\,\omega+6\,d\,q_4^2\,\omega\big]+q_4\,\big[
c\,d^2-d^3+2\,d\,(5+2\,c+2\,q_4^2\,\omega)+4\,(6\,q_4^2\,\omega-7)\big]\Big],\\
Q_4=q_4^3\,\Big[8\,k^2\,q_4\,\omega+q_4\,\big[
2\,d\,q_4\,\omega+8\,q_4\,\omega+c\,(d^2+2\,d-2)+d-8\big]\\
\,\,\,\,\,\,\,\,\,\,\,\,\,\,\,\,\,\,\,\,\,\,\,\,\,\,\,\,\,\,\,\,\,\,
\,\,\,\,\,\,\,\,\,\,\,\,\,\,\,\,\,\,\,\,\,\,\,\,\,\,
+k\,\big[
2\,d\,q_4^2\,\omega+16\,q_4^2\,\omega+c\,(3\,d-2)+3\,d-10\big]\Big].
\end{array}
\right.
\end{equation}
The equation (\ref{u49-3}) leads to $Q_i\,=\,0$ or all $i=0,1,2,3,4$. By solving these equation we can get two cases:\\

\textbf{Case (3.1.1):}  When $q_4\,\neq\,-k$, we can obtain the following solution $q_4=-\dfrac{k}{2}$ and $d=c=4$. Using (\ref{u48-3}), (\ref{u47-3}), (\ref{u46-3}) and the transformation (\ref{u42-3}), we have the following solution of the field equation
\begin{equation}\label{u411-3}
\left\{
\begin{array}{ll}
E(z,t)=q_1\,(z+a)^{\dfrac{k^2\,\omega}{4}-\dfrac{1}{2}}\,(t+b)^{\dfrac{9}{2}-\dfrac{k^2\,\omega}{4}}\,
\Big[\Big(\dfrac{z+a}{t+b}\Big)^2-1\Big]^{3-\dfrac{k^2\,\omega}{4}}\,\Big[\Big(\dfrac{z+a}{t+b}\Big)^2+1\Big]^{-\dfrac{1}{2}},\\
\\
G(z,t)=q_2\,(z+a)\,(t+b)^3\,\Big[\Big(\dfrac{z+a}{t+b}\Big)^2+1\Big],\\
\\
\phi(z,t)=q_3\,(z+a)^{k/2}\,(t+b)^{k/2}.
\end{array}
\right.
\end{equation}
The metric of this corresponding solution can be written in the following form:
\begin{equation}  \label{s311}
\begin{array}{ll}
ds_{311}^2=q_1\,Z^{\dfrac{k^2\,\omega}{4}-\dfrac{1}{2}}\,T^{\dfrac{9}{2}-\dfrac{k^2\,\omega}{4}}\,
\Big[\Big(\dfrac{Z}{T}\Big)^2-1\Big]^{3-\dfrac{k^2\,\omega}{4}}\,\Big[\Big(\dfrac{Z}{T}\Big)^2+1\Big]^{-\dfrac{1}{2}}\,\big(dZ^2-dT^2\big)\\
\\
\,\,\,\,\,\,\,\,\,\,\,\,\,\,\,\,\,\,\,\,\,\,\,\,\,\,\,\,\,\,\,\,\,\,\,\,\,\,\,\,\,\,\,\,\,\,\,\,\,\,
+q_2\,Z\,T\,(Z^2+T^2)\,\big(dx^2+dy^2\big),
\end{array}
\end{equation}
where $Z=z+a$ and $T=t+b$ while $q_1$, $q_2$, $a$, $b$, $k$ and $\omega$ are an arbitrary constants. The energy density and pressure for the model (\ref{s311}) are given by
\begin{equation}  \label{u412-3}
\begin{array}{ll}
\rho(z,t)\,=\,p(z,t)\,=\,0.
\end{array}
\end{equation}

\textbf{Case (3.1.2):}  When $q_4\,=\,-k$, we can obtain the following two solutions:\\

\textbf{Case (3.1.2.1):} $d=1$. Using (\ref{u48-3}), (\ref{u47-3}), (\ref{u46-3}) and the transformation (\ref{u42-3}), we have the following solution of the field equation
\begin{equation}\label{u413-3}
\left\{
\begin{array}{ll}
E(z,t)=q_1\,(t+b)^{c},\\
\\
G(z,t)=q_2\,(t+b),\\
\\
\phi(z,t)=q_3\,(t+b)^{k}.
\end{array}
\right.
\end{equation}
The metric of this corresponding solution can be written in the following form:
\begin{equation}  \label{s3121}
\begin{array}{ll}
ds_{3121}^2=q_1\,(t+b)^{c}\,\big(dz^2-dt^2\big)+q_2\,(t+b)\,\big(dx^2+dy^2\big),
\end{array}
\end{equation}
where $q_1$, $q_2$, $b$, $c$, $k$ and $\omega$ are an arbitrary constants. The energy density and pressure for the model (\ref{s3121}) are given by
\begin{equation}  \label{u414-3}
\begin{array}{ll}
\rho(z,t)\,=\,p(z,t)\,=\,\Big(\dfrac{1+2\,c-2\,k^2\,\omega}{4\,q_1\,K}\Big)(t+b)^{-2-c}.
\end{array}
\end{equation}

\textbf{Case (3.1.2.2):} $d=\dfrac{2}{c}$. Using (\ref{u48-3}), (\ref{u47-3}), (\ref{u46-3}) and the transformation (\ref{u42-3}), we have the following solution of the field equation
\begin{equation}\label{u415-3}
\left\{
\begin{array}{ll}
E(z,t)=q_1\,(z+a)^{\dfrac{c}{2}-1}\,(t+b)^{\dfrac{c}{2}+1},\\
\\
G(z,t)=q_2\,(z+a)^{\dfrac{2}{c}-1}\,(t+b),\\
\\
\phi(z,t)=q_3\,(t+b)^{k}.
\end{array}
\right.
\end{equation}
The metric of this corresponding solution can be written in the following form:
\begin{equation}  \label{s3122}
\begin{array}{ll}
ds_{3122}^2=q_1\,Z^{\dfrac{c}{2}-1}\,T^{\dfrac{c}{2}+1}\,\big(dZ^2-dT^2\big)+q_2\,Z^{\dfrac{2}{c}-1}\,T\,\big(dx^2+dy^2\big),
\end{array}
\end{equation}
where $Z=z+a$ and $T=t+b$ while $q_1$, $q_2$, $a$, $b$, $c$, $k$ and $\omega$ are an arbitrary constants. The energy density and pressure for the model (\ref{s3122}) are given by
\begin{equation}  \label{u415-3}
\begin{array}{ll}
\rho(z,t)\,=\,\dfrac{1}{4\,q_1\,c^2\,K}\Big[c^2(3+c-2\,k^2\,\omega)\,Z^2-(c-1)\,(c-2)\,(c+6)\,T^2\Big]\,Z^{-1-\dfrac{c}{2}}\,T^{-3-\dfrac{c}{2}},
\end{array}
\end{equation}

\begin{equation}  \label{u416-3}
\begin{array}{ll}
p(z,t)\,=\,\dfrac{1}{4\,q_1\,c^2\,K}\Big[c^2(3+c-2\,k^2\,\omega)\,Z^2-(c-1)\,(c-2)^2\,T^2\Big]\,Z^{-1-\dfrac{c}{2}}\,T^{-3-\dfrac{c}{2}},
\end{array}
\end{equation}
The spatial volume $V$ is given by
\begin{equation}  \label{u415-1-3}
\begin{array}{ll}
V(z,t)=q_1\,q_2\,Z^{\dfrac{c^2-4\,c+4}{2c}}\,T^{\dfrac{c}{2}+2}.
\end{array}
\end{equation}
The scalar expansion $\Theta$ and the shear scalar $\sigma$ are given:
\begin{equation}  \label{u416-1-3}
\begin{array}{ll}
\Theta=\Big(\dfrac{c+6}{4\,\sqrt{q_1}}\Big)\,Z^{-\dfrac{3}{2}-\dfrac{c}{4}}\,T^{\dfrac{1}{2}-\dfrac{c}{4}},
\end{array}
\end{equation}

\begin{equation}  \label{u417-1-3}
\begin{array}{ll}
\sigma=\dfrac{c^2}{48\,q_1}\,Z^{-3-\dfrac{c}{2}}\,T^{1-\dfrac{c}{2}}.
\end{array}
\end{equation}
The deceleration parameter $\mathbf{q}$ is given by:
\begin{equation}  \label{u418-1-3}
\begin{array}{ll}
\mathbf{q}=\dfrac{(c+6)^4}{128\,q_1^2}\,Z^{2-c}\,T^{-6-c}.
\end{array}
\end{equation}

\textbf{Case (3.2):}  When $d\,=\,2$, then
\begin{equation}\label{u417-3}
\begin{array}{ll}
\Omega(\xi)=\dfrac{q_2}{\xi},
\end{array}
\end{equation}

\begin{equation}\label{u418-3}
\begin{array}{ll}
\Psi(\xi)=q_1\,\xi^{\alpha_1}\,(\xi^2-1)^{q_5},
\end{array}
\end{equation}
where $q_1$, $q_2$ and $q_5$ are arbitrary constants while $\alpha_1=(3\,q_4+k)\,(q_4+k)\,\omega-\dfrac{3}{2}$ and $c=1-2\,q_4\,\omega\,(q_4+k)$.  Therefore, using (\ref{u48-3}), (\ref{u47-3}), (\ref{u46-3}) and the transformation (\ref{u42-3}), we have the following solution of the field equation
\begin{equation}\label{u419-3}
\left\{
\begin{array}{ll}
E(z,t)=q_1\,(z+a)^{c}\,\Big(\dfrac{z+a}{t+b}\Big)^{\alpha_1}\,\Big[\dfrac{z+a}{t+b}-1\Big]^{q_5},\\
\\
G(z,t)=q_2\,(z+a)\,(t+b),\\
\\
\phi(z,t)=q_3\,(z+a)^{q_4+k}\,(t+b)^{-q_4}.
\end{array}
\right.
\end{equation}
The metric of this corresponding solution can be written in the following form:
\begin{equation}  \label{s32}
\begin{array}{ll}
ds_{32}^2=q_1\,Z^{c}\,\Big(\dfrac{Z}{T}\Big)^{\alpha_1}\,\Big[\dfrac{Z}{T}-1\Big]^{q_5}\,\big(dZ^2-dT^2\big)
+q_2\,Z\,T\,\big(dx^2+dy^2\big),
\end{array}
\end{equation}
where $Z=z+a$ and $T=t+b$ while $q_1$, $q_2$, $q_5$, $a$, $b$, $k$ and $\omega$ are an arbitrary constants. The energy density and pressure for the model (\ref{s32}) are given by
\begin{equation}  \label{u420-3}
\begin{array}{ll}
\rho(z,t)=p(z,t)=\Big(\dfrac{2\,(1-q_5)-(2\,q_4+k)^2\,\omega}{2\,q_1\,K}\Big)\,\Big(\dfrac{Z^2}{T^2}-1\Big)^{-q_5}\,Z^{c/2}\,T^{\alpha_1-2}.
\end{array}
\end{equation}
The spatial volume $V$ is given by
\begin{equation}  \label{u415-1-4}
\begin{array}{ll}
V(z,t)=q_1\,q_2\,Z^{1+c+\alpha_1}\,T^{1-\alpha_1}\,\Big(\dfrac{Z^2}{T^2}-1\Big)^{q_5}.
\end{array}
\end{equation}
The scalar expansion $\Theta$, the shear scalar $\sigma$ and the deceleration parameter $\mathbf{q}$ are given:
\begin{equation}  \label{u416-1-4}
\begin{array}{ll}
\Theta=\Big(\dfrac{(\alpha_1-2+2\,q_5)\,Z^2+(\alpha_1-2)\,T^2}{2\,\sqrt{q_1}\,(T^2-Z^2)}\Big)\,Z^{-\dfrac{\alpha_1-c}{2}}\,T^{\dfrac{\alpha_1}{2}-1},
\end{array}
\end{equation}

\begin{equation}  \label{u417-1-3}
\begin{array}{ll}
\sigma=\Big(\dfrac{(\alpha_1+1)\,T^2-(\alpha_1+1+2\,q_5)\,Z^2}{12\,q_1\,(T^2-Z^2)^2}\Big)\,Z^{-\alpha_1-c}\,T^{\alpha_1-2},
\end{array}
\end{equation}

\begin{equation}  \label{u418-1-3}
\begin{array}{ll}
\mathbf{q}=\dfrac{1}{8\,q_1^2\,(T^2-Z^2)^4}\,\Big[(\alpha_1-2+2\,q_5)^2\,Z^4-2\,\big[2\,q_5\,(\alpha_1-5)+(\alpha_1-2)^2\big]\,T^2\,Z^2\\
\,\,\,\,\,\,\,\,\,\,\,\,\,\,\,\,\,\,\,\,\,\,\,\,\,
+(\alpha_1-2)^2\,Z^4\Big]\,\Big[(\alpha_1-2)\,T^2-(\alpha_1-2+2\,q_5)^2\,Z^2\Big]\,Z^{-2\,(\alpha_1+c)}\,T^{2\,(\alpha_1-2)}.
\end{array}
\end{equation}

\textbf{Case (4):} When $n\,\neq\,-2$ and $c_1\,\neq\,0$, the similarity variable and similarity functions can be written as the following:
\begin{equation}\label{u42-4}
\left\{
\begin{array}{ll}
\xi\,=\,\dfrac{z+a}{t+b},\\
E(z,t)\,=\,(z+a)^c\,\Psi(\xi),\\
G(z,t)\,=\,(z+a)^d\,\Omega(\xi),\\
\phi(z,t)\,=\Big(\Phi(\xi)+\tilde{k}\,\ln[z+a]\Big)^{\dfrac{2}{n+2}},
\end{array}
\right.
\end{equation}
where $a=\dfrac{c_2}{c_1}$, $b=\dfrac{c_3}{c_1}$, $c=\dfrac{c_4}{c_1}$, $d=\dfrac{c_5}{c_1}$ and $\tilde{k}=\dfrac{(n+2)\,c_6}{2\,c_1}$ are an arbitrary constants. Substituting the transformations (\ref{u42-4}) in the field Eqs. (\ref{u26})-(\ref{u28}) lead
to the following system of ordinary differential equations:

\begin{equation}\label{u43-4}
\begin{array}{ll}
\xi^2\,(\xi^2-1)\,\Big(\dfrac{\Psi\,\Psi''-\Psi'^2}{\Psi^2}\Big)
-\xi^2\,(\xi^2+1)\,\Big(\dfrac{\Psi\,\Omega''-\Psi'\,\Omega'}{\Psi\,\Omega}\Big)+\xi^2\,\Big(\dfrac{\Omega'^2}{\Omega^2}\Big)\\
\\
\,\,\,\,\,\,\,\,\,\,\,\,\,\,\,\,\,\,\,\,\,\,\,\,\,
+\xi\,(d+2\,\xi^2)\,\dfrac{\Psi'}{\Psi}+\xi\,(c-2\,\xi^2)\,\dfrac{\Omega'}{\Omega}
=\dfrac{8\,\omega}{(n+2)^2}\,\,\Big(\xi\,\Phi'+\tilde{k}\Big)^2-c\,d-d-c,
\end{array}
\end{equation}

\begin{equation}\label{u44-4}
\begin{array}{ll}
\xi\,\Big[\dfrac{\Omega''}{\Omega}-\dfrac{\Omega'^2}{2\,\Omega^2}-\dfrac{\Psi'\,\Omega'}{\Psi\,\Omega}+\dfrac{4\,\omega\,\Phi'^2}{(n+2)^2}\Big]
=\dfrac{d\,\Psi'}{2\,\Psi}
+\dfrac{(c-d-2)\,\Omega'}{2\,\Omega}-\dfrac{4\,\tilde{k}\,\omega\,\Phi'}{(n+2)^2},
\end{array}
\end{equation}

\begin{figure}
\begin{center}
\includegraphics[width=4.5in]{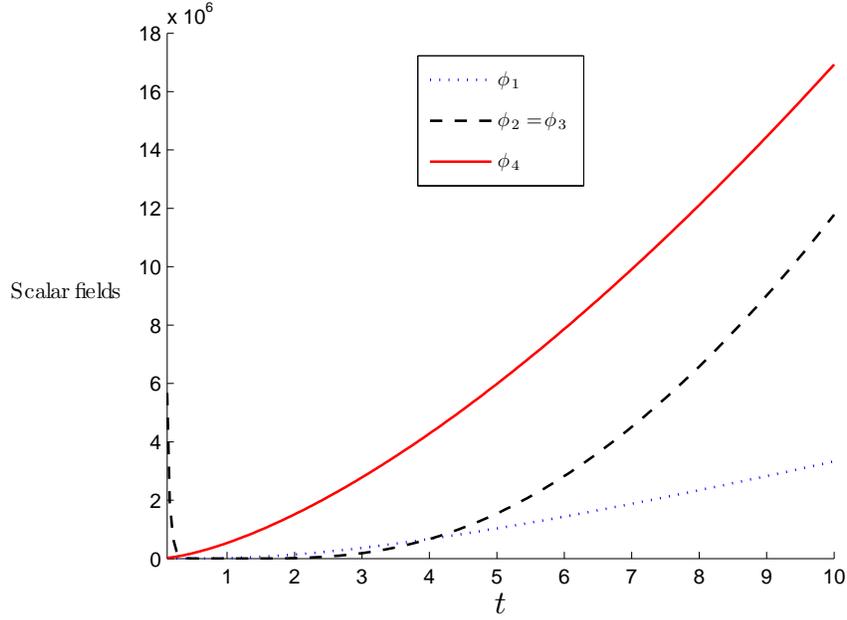}
\caption{The plot of scalar fields $(\phi's)$ versus time (t).}
\label{fg:ahmadst01F1.eps}
\end{center}
\end{figure}

\begin{equation}\label{u45-4}
\begin{array}{ll}
\xi^2\,(1-\xi^2)\,\Big[\dfrac{\Phi''}{\Phi}+\dfrac{\Omega'\,\Phi'}{\Omega\,\Phi}\Big]
=\dfrac{1}{\Phi}\,\Big[\xi\,(2\,\xi^2-d)\,\Phi'-\tilde{k}\,\xi\,\Big(\dfrac{\Omega'}{\Omega}\Big)-\tilde{k}\,(d-1)\Big].
\end{array}
\end{equation}

If one solves the system of second order non-linear ordinary differential equations (\ref{u43-4})-(\ref{u45-4}), he can obtain the exact solutions of the original field equations (\ref{u26})-(\ref{u28}) corresponding to reduction (\ref{u42-4}). The system (\ref{u43-4})-(\ref{u45-4}) is very difficult to solve in general form. This system may be solved in some special cases as the following:\\

We assume the solution of the function $\Phi(\xi)$ in the form:
\begin{equation}\label{u46-4}
\begin{array}{ll}
\Phi(\xi)=\ln[q_3\,\xi^{q_4}],
\end{array}
\end{equation}
where $q_3$ and $q_4$ are arbitrary non-zero constants.
Here, we deduce the following solutions:\\

\textbf{Solution (4.1):} The metric takes the form (\ref{s311}) with the scalar field
\begin{equation}\label{u47-4}
\begin{array}{ll}
\phi_{1}^{\frac{n}{2}+1}=\ln\Bigg[q_3\,\Big(Z\,T\Big)^{\dfrac{k\,(n+2)}{4}}\Bigg],
\end{array}
\end{equation}
where $Z=z+a$, $T=t+b$ while $q_3$, $k$, $a$, $b$ and $n$ are arbitrary constants.\\

\textbf{Solution (4.2):} The metric takes the form (\ref{s3121}) with the scalar field
\begin{equation}\label{u48-4}
\begin{array}{ll}
\phi_{2}^{\frac{n}{2}+1}=\ln\Bigg[q_3\,T^{\dfrac{k\,(n+2)}{2}}\Bigg],
\end{array}
\end{equation}
where $T=t+b$ while $q_3$, $k$, $b$ and $n$ are arbitrary constants.\\

\textbf{Solution (4.3):} The metric takes the form (\ref{s3122}) with the scalar field
\begin{equation}\label{u49-4}
\begin{array}{ll}
\phi_{3}^{\frac{n}{2}+1}=\ln\Bigg[q_3\,T^{\dfrac{k\,(n+2)}{2}}\Bigg],
\end{array}
\end{equation}
where $T=t+b$ while $q_3$, $k$, $b$ and $n$ are arbitrary constants.\\

\textbf{Solution (4.4):} The metric takes the form (\ref{s32}) with the scalar field
\begin{equation}\label{u410-4}
\begin{array}{ll}
\phi_{4}^{\frac{n}{2}+1}=\ln\Bigg[q_3\,Z^{q_4+\dfrac{k\,(n+2)}{2}}\,T^{-q_4}\Bigg],
\end{array}
\end{equation}
where $Z=z+a$, $T=t+b$ while $q_3$, $q_4$, $k$, $a$, $b$ and $n$ are arbitrary
constants.

\section{Conclusion}
In this paper, we have studied the plane symmetric inhomogeneous cosmological models with
perfect fluid as source of matter within the framework of scalar-tensor theory of gravitation.
The Lie group analysis method transforms the Einstein field equations into the system of ordinary
differential equations. We obtained a new class of exact solutions of field equations for the models under
consideration by using symmetric group analysis method. The derived models are singular in nature and it
has big bang singularity at $t = -b$, except the model (\ref{s311}). The spatial volume is zero
whereas all the physical parameters $\rho$, $p$ and $\sigma$ assume infinite value at initial moment $t = -b$.
$\rho$ and $p$ are decreasing function of time for model (\ref{s3121}), (\ref{s3122}) and (\ref{s32})
while it has zero value for
model (\ref{s311}). Therefore, in our analysis, the singularity free
model (\ref{s311}) resembles with dusty universe.
For all derived models, the scalar functions have similar nature which is depicted in Figure 1.



\begin{thebibliography}{99}

\bibitem{ali1} Ali, A.T.: Phys. Scr. \textbf{79(3)}, 035006 (2009)

\bibitem{ali2} Ali, A.T.: Phys Scr \textbf{87(1)}, 015002 (2013)

\bibitem{yadav2} Ali, A. T., Yadav, A. K.: arXiv: 1305.4631 [gr-qc]

\bibitem{anguige1} Anguige, K.: Class Quantum Gravi \textbf{17}, 2117 (2000)

\bibitem{ben1} Bennet, C.L., et al.: Astrophys. J. Suppl. Ser. \textbf{148}, 1 (2003)

\bibitem{brans1} Brans, C., Dicke, R.H.: Physical Review \textbf{124}, 925 (1961)

\bibitem{dasilva1} Da Silva, M.F.A., Wang, A.: Phys Lett A \textbf{A244}, 462 (1998)

\bibitem{jamil1} Jamil, M., Ali, S., Momeni, D., Myrzakulov, R.: Europeon Physical Journal C \textbf{72}, 1998 (2012)

\bibitem{elsa2} El-Sabbagh, M.F., Ali, A.T.: Commun Nonlinear Sci Numer Simulat \textbf{13} 1758 (2008) 

\bibitem{marra1} Marra, V., Paakkonen, M.: JCAP \textbf{01}, 025 (2008) 

\bibitem{nouri1} Nouri-Zonoz, M.,Tavanfar, A.R.: Class Quantum Gravi \textbf{18}, 4293 (2001) 

\bibitem{pradhan1} Pradhan, A., Pandey, H.R.: Int J Mod Phys. D \textbf{12}, 941 (2003)

\bibitem{pradhan2} Pradhan, A., Rai, K. K., Yadav, A. K.: Braz. J. Phys. \textbf{37}, 1084 (2007)

\bibitem{perl1} Perlmutter, S., et. al.:Astrophys. J \textbf{517}, 565 (1999)

\bibitem{reddy1} Reddy, D.R.K., Subba, R.M.V., Koteswara, R.G.: Astrophys Space Sci \textbf{306}, 171 (2006)

\bibitem{ries1} Riess, A. G., et al.: Astron J. \textbf{116}, 1009 (1998)

\bibitem{saez1} Saez, D., Ballester, V.J.: Phys Lett A \textbf{113}, 467 (1985)

\bibitem{singh1} Singh, T., Agarwal, A.K.: Astrophys Space Sci \textbf{182}, 289 (1991)

\bibitem{socco1} Socorro, J., Sabido, M., Sanchez, M. A., Frias palos, M. G.: 
Revista Mexicana de Fisica, \textbf{56}, 166 (2010)

\bibitem{yadav1} Yadav, A. K.: Int. J. Theor. Phys. \textbf{49}, 1140 (2011)


\end{thebibliography}
\end{document}